\documentclass[conference]{IEEEtran}
\IEEEoverridecommandlockouts
\usepackage{cite}
\usepackage{amsmath,amssymb,amsfonts}
\usepackage{algorithmic}
\usepackage{graphicx}
\usepackage{textcomp}
\usepackage{xcolor}
\usepackage{multirow}
\usepackage{enumitem}
\usepackage{booktabs}
\usepackage{bbm}
\usepackage{subfigure}
\def\BibTeX{{\rm B\kern-.05em{\sc i\kern-.025em b}\kern-.08em
    T\kern-.1667em\lower.7ex\hbox{E}\kern-.125emX}}
\begin{document}

\title{DuoServe-MoE: Dual-Phase Expert Prefetch and Caching for LLM Inference QoS Assurance\\
%\title{DuoServe-MoE: Dual-Phase Expert Prefetch and Cache Scheduling for Efficient MoE LLM Inference\\
% {\footnotesize \textsuperscript{*}Note: Sub-titles are not captured for https://ieeexplore.ieee.org  and
}
\author{
\IEEEauthorblockN{Yuning Zhang, Grant Pinkert, Nan Yang, Yanli Li, and Dong Yuan}
\IEEEauthorblockA{\textit{School of Electrical and Computer Engineering, The University of Sydney}\\
Darlington, NSW 2008, Australia
}
\thanks{Corresponding author: Dong Yuan.}
}
% \and
% \and
% \and
% \and
% \and

\maketitle

\begin{abstract}
Large Language Models (LLMs) are increasingly deployed as Internet/Web services (LLM-as-a-Service) with strict latency Service-Level Objectives (SLOs) under tight GPU memory budgets. Mixture-of-Experts (MoE) models improve quality and throughput via sparse expert activation, but serving them efficiently is challenging because expert weights dominate memory footprint and incur costly host--device transfers when offloaded. Moreover, MoE serving exhibits a phase disparity: the prefill phase tends to activate experts densely across many tokens, while the decode phase activates only a few experts per step. A uniform expert loading/caching policy across phases leads to either peak-memory blowup (prefill) or tail-latency inflation (decode). We present DuoServe-MoE, a QoS-oriented MoE serving system that decouples prefill and decode and applies phase-specialized expert scheduling. For prefill, DuoServe-MoE uses a two-stream CUDA pipeline to overlap expert prefetching with non-MoE computation, reducing expert residency time and peak GPU memory. For decode, it employs a lightweight layer-level predictor trained offline from activation traces to prefetch only likely experts without model changes. Experiments on representative MoE LLMs show that DuoServe-MoE improves TTFT by up to $5.34\times$ and end-to-end latency by up to $7.55\times$ over representative baselines, while maintaining low runtime GPU memory usage under resource-constrained deployment.
\end{abstract}

\section{Introduction}
Large Language Models (LLMs) and their vast parameter scales have achieved remarkable performance across a wide range of tasks\cite{zhao2023survey}, and are increasingly delivered as Internet/Web-based services, e.g., conversational assistants and enterprise model APIs\cite{openai2024chatgpt}. In these deployments, service providers must satisfy strict latency Service-Level Objectives (SLOs) and control the infrastructure cost under limited GPU memory, especially on cost-sensitive single-GPU servers and edge environments\cite{qu2024mobile,dhar2024empirical}. Such deployments are common in practical LLM-as-a-Service settings, where service instances are often provisioned with only one commodity GPU and must still provide responsive inference under fluctuating request characteristics. As a result, Quality-of-Service (QoS) management, particularly reducing end-to-end (E2E) latency, controlling tail latency, and preventing peak-memory blowups, has become a first-class design goal for LLM serving systems.

Mixture of Experts (MoE) models\cite{shazeer2017} are a prominent approach to scale model capacity by replacing the feed-forward network (FFN) block in Transformers\cite{vaswani2017attention} with a set of experts and using a gate to route each token to a small subset of experts. This sparse activation enables high model capacity without proportionally increasing per-token computation, and has been adopted by large-scale LLMs such as Switch Transformer, Mixtral, and DeepSeek\cite{fedus2022switch,jiang2024mixtral,dai2024deepseekmoe}. However, from a service-computing perspective, MoE introduces a new bottleneck: expert weights dominate the memory footprint, and moving experts between CPU memory and GPU memory can inflate latency, increase response-time variance, and degrade service stability.

To fit MoE models into limited GPU memory, one common strategy is to keep only a subset of experts resident in GPU memory and offload the rest to CPU memory, effectively turning expert weights into a managed service resource. Prior works reduce memory cost with multi-level expert caching\cite{yu2024moesys} or offload experts on demand after gating (e.g., Huggingface Accelerate\cite{huggingface2022accelerate}). However, on-demand fetching places host--device transfers on the critical path and can significantly increase E2E latency. Prefetch-based approaches\cite{hwang2024pre,yi2023edgemoe} attempt to predict the potential experts before the gate and prefetch them to hide transfer latency, but accurate prediction becomes increasingly difficult as modern MoE LLMs scale up and adopt Top-$k$ routing, where each token may activate multiple experts. In service settings, this difficulty directly translates into unstable latency behavior, since prediction misses trigger extra transfers and delay request completion.

Existing MoE serving systems often assume more regular expert routing to simplify prediction and caching, which does not fully reflect the more complex Top-$k$ activation behavior of modern MoE LLMs~\cite{xue2024moe, gavhane2025moebeyondlearningbasedexpertactivation}. Some approaches further rely on model fine-tuning or architectural modification~\cite{du2024sida,hwang2024pre,raje2026melinoefinetuningenablesmemoryefficient}, increasing deployment cost and reducing practicality for QoS-oriented single-GPU serving. More importantly, MoE serving is inherently \emph{phase-dependent}. Autoregressive inference consists of a \emph{prefill} phase (processing many prompt tokens in parallel) and a \emph{decode} phase (generating one token per step). During prefill, expert activation becomes effectively dense across tokens, which pushes systems to load many experts to maximize parallelism (e.g., vLLM\cite{kwon2023efficient} and DeepSpeed\cite{aminabadi2022deepspeed}), but this can cause peak-memory blowups and Out-of-Memory (OOM) failures on small GPUs. During decode, expert activation returns to sparse routing, and the dominant QoS risk becomes transfer stalls that inflate per-step latency and tail latency. Therefore, prefill and decode stress different service bottlenecks: prefill is more sensitive to memory pressure and transient residency, whereas decode is more sensitive to fine-grained transfer delay and latency stability. These observations motivate a QoS-oriented MoE service runtime that treats prefill and decode differently, rather than applying a single expert scheduling policy across the entire request. However, existing MoE inference serving systems have largely overlooked this phase-specific optimization opportunity~\cite{yu2025taminglatencymemorytradeoffmoebased, chen2025ktransformers}.

To address the challenges of expert prediction and the distinct activation patterns between the prefill and decode phases in Mixture-of-Experts (MoE) inference, we design an inference serving system called DuoServe-MoE, which applies dedicated expert scheduling strategies for each phase. DuoServe-MoE is designed for resource-constrained single-GPU service deployments, where maintaining responsive latency under limited memory is more important than maximizing cluster-scale throughput. In summary, our paper presents the following contributions:

\begin{itemize}[leftmargin=*]
\item Considering the two-phase inference characteristics of LLMs, we developed two distinct expert scheduling strategies. In the prefill stage, we employ a two-level pipelining approach to parallelize communication and computation. During the decoding stage, we use a lightweight prediction model to prefetch the experts that will be activated into GPU memory, reducing communication overhead. To our knowledge, DuoServe-MoE is the first system that explicitly separates expert scheduling between prefill and decode for resource-constrained single-GPU MoE serving.

\item We design a layer-level prediction method to forecast expert selection for each token without changing the model architecture or accuracy. We incorporated the popularity of experts at each layer, the inter-layer experts' affinity, and the expert activation paths into this learning-based prediction model. Popularity refers to how frequently a particular expert is selected across inference runs, while affinity measures the likelihood of specific experts being activated in consecutive layers. We abstracted the combination of multiple experts per layer into a single expert's influence on the selection of experts in the subsequent layer, thereby simplifying the training process while maintaining notable prediction accuracy. This lightweight model ensures that all processes of DuoServe-MoE can be completed on the same GPU, alleviating the pressure of resource constraints. 

\item Overall, we propose DuoServe-MoE, an inference serving system for MoE-based LLMs that reduces GPU memory pressure by offloading expert model weights to CPU memory and applying phase-specific expert scheduling for the prefill and decode stages. We evaluate DuoServe-MoE on four representative MoE models using a single GPU. Compared to the baselines, DuoServe-MoE achieves 1.78$\times$ to 5.34$\times$ improvement in TTFT and 1.42$\times$ to 7.55$\times$ improvement in end-to-end latency, while keeping peak runtime GPU memory usage low under offloading-based deployment. 
\end{itemize}
\section{Background and Motivation}

\subsection{Mixture of Experts and Activation Mechanism} \label{2.1}
\textbf{Mixture of Experts Architecture.} Mixture of Experts (MoE)\cite{fedus2022switch} is a model architecture originally introduced to tackle the challenges of efficient model parameter utilization and increased representational power. The core idea behind MoE is to partition the workload across multiple specialized sub-models, known as “experts.” The output is expressed as a weighted sum of the outputs from experts. Each expert produces an output based on the input, and the gate function determines the weight assigned to each expert. The gate function values are non-negative and sum to 1, representing the contribution of each expert to the final output. The gate function is typically modeled using a separate neural network. Gshard\cite{lepikhin2021gshard} employs a softmax function to compute the weights based on the input. In current MoE LLMs, the gate usually performs Top-$k$ routing over model-specific expert pools, and some architectures further include shared experts that contribute to all tokens or are merged outside the regular routed-expert path.

\begin{figure}[t]
  \centering
  \includegraphics[width=0.38\textwidth]{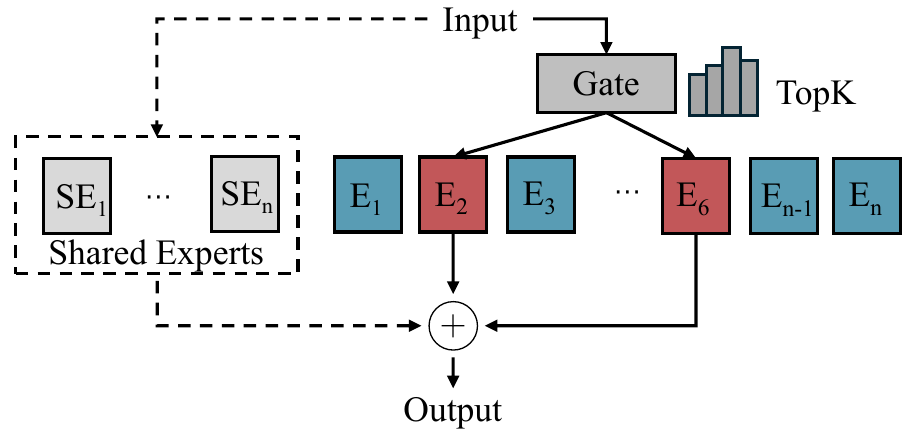} 
  \caption{Unified MoE architecture illustration. The gate selects Top-$k$ experts from $n$ routed experts (model-dependent), and some models additionally use shared experts (dashed) that are fused into the final output.}
  \label{Moe_arch}
\end{figure}

\textbf{Complexity of MoE Architecture in LLMs.} Figure \ref{Moe_arch} presents a unified MoE-layer view used by modern LLMs. Given an input token, the gate selects Top-$k$ experts from an expert pool of size $n$, where both $n$ and $k$ are model-dependent (e.g., Top-1, Top-2, or higher). Some MoE variants additionally include shared experts that are not exclusively routed by the token-level Top-$k$ decision; these optional components are shown with dashed connections in Fig.~\ref{Moe_arch}. This heterogeneity in expert count, activated-expert count, and shared-expert design increases the complexity of expert offloading and cache scheduling in practical serving systems.

\begin{figure}[]
    \centering
    \subfigure[The popularity of experts in each layer.]{
        \includegraphics[width=0.45\linewidth]{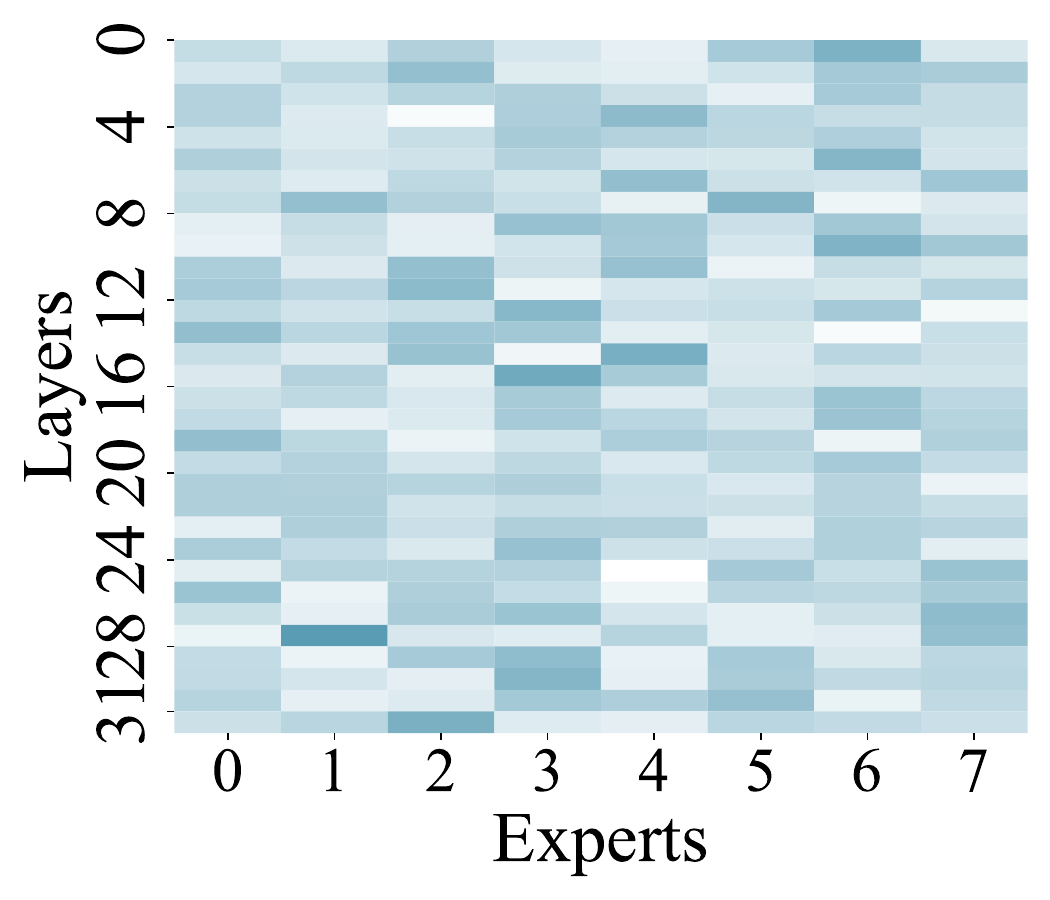}
        \label{fig:sub1}
    }
    \subfigure[The inter-layer affinity between layer 0 and layer 1.]{
        \includegraphics[width=0.45\linewidth]{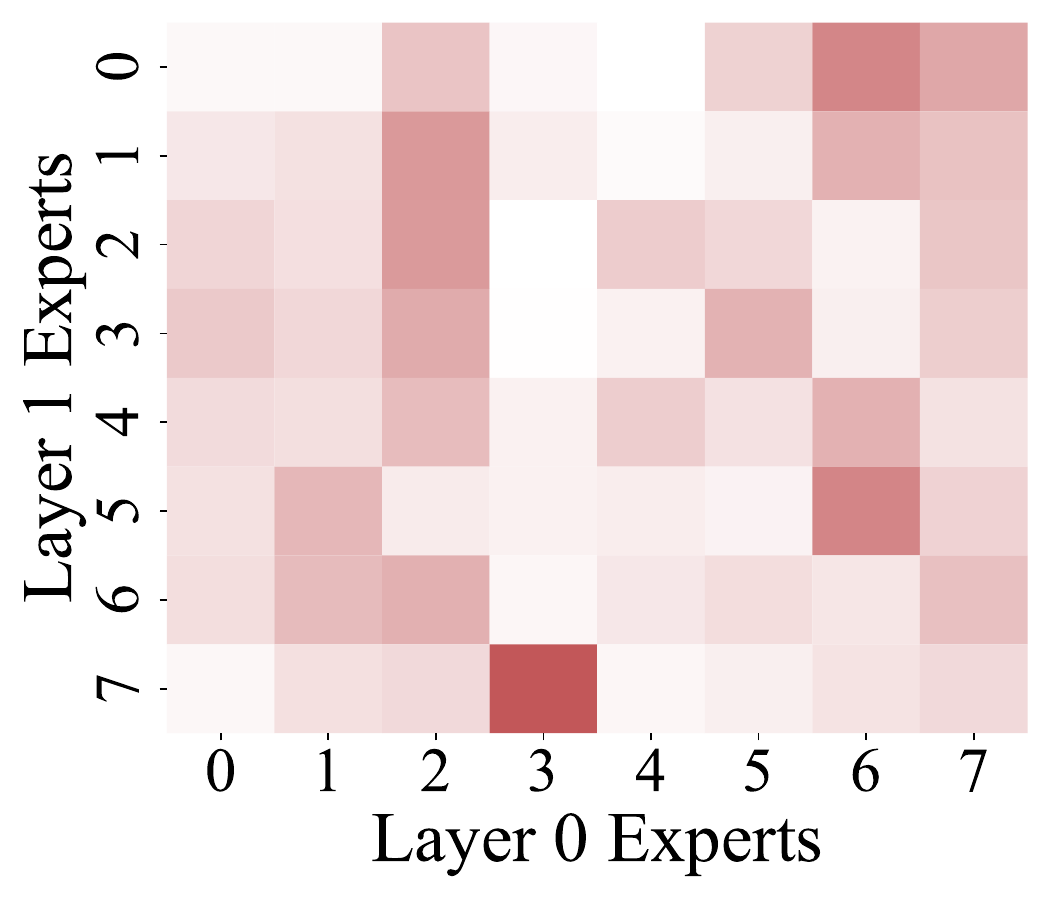}
        \label{fig:sub2}
    }
    \caption{Popularity and Affinity in MoE-based LLMs. Darker colors represent higher probabilities.}
    \label{fig:side_by_side}
\end{figure}

\textbf{Popularity and Affinity in Experts Activation.} 
Although the activation mechanism of MoE in LLMs is becoming complex, it is not entirely random. Figure \ref{fig:side_by_side} shows the popularity of different expert models across different layers in Mixtral, as well as the affinity between different experts in layer 0 and layer 1. 
Yao et al.\cite{yao2024exploiting} discovered an inter-layer affinity pattern in models employing a Top-1 mechanism: when an expert is activated in one layer, specific experts in the subsequent layer are more likely to be activated. This inter-layer affinity still exists in models using Top-2 or higher expert selection mechanisms. Even with multiple experts being activated per layer, there is a recognizable pattern where certain experts are more likely to be activated in subsequent layers based on previous activations. This demonstrates that certain experts are more frequently activated and that there are discernible relationships between experts in consecutive layers.

\textbf{Challenge \#1: Accurately predict the expert selection without changing the LLMs.} These expert activation characteristics create opportunities for designing MoE inference serving systems under resource-constrained scenarios. By predicting which experts will be activated at each layer and prefetching activated model weights onto the GPU, memory consumption and data transmission overhead can be reduced to enhance inference speed. Therefore, predicting expert activations accurately becomes a key focus. However, as shown in Figure \ref{fig:side_by_side}, while there are discernible activation patterns based on popularity and affinity, the distribution is not highly concentrated. Directly designing a heuristic algorithm based solely on these patterns would not achieve high accuracy. Instead, learning-based methods are often preferred as they allow the model to learn activation routes more accurately. However, some prior works fine-tune the LLMs, which affects the model accuracy and raises a challenge for the resource-constrained environment.
\subsection{MoE in Large Language Models}
\textbf{Prefill and Decode.} In the inference process of LLMs, there are two phases: prefill and decode. During the prefill stage, the model receives the complete input sequence and processes the parallel computation of these tokens. This means that all input tokens can be processed simultaneously during this stage, rather than decoding each token step by step. In the decode stage, the model generates each output token sequentially, using the partially generated output (along with cached features from the prefill stage) to generate new tokens one by one. At this stage, the model uses the already generated tokens and context information to infer the next token, making the process inherently sequential and unsuitable for parallelization like in the prefill stage.

\textbf{Activation Differences in LLMs.} For a single token, expert activation in an MoE layer is sparse. However, during prefill, an LLM processes many tokens in parallel, and the union of token-level Top-$k$ routing decisions can become effectively dense. For instance, Mixtral can activate all experts in a layer during prefill when routes are aggregated across tokens in one batch. As a result, the number of activated experts in prefill can be considerably larger than in decode, leading to significant memory demand and challenges in managing parallel activations efficiently. Unlike the prefill stage, each iteration processes only one token during decode, so expert activation remains sparse. \textbf{However, in single-GPU deployments without expert parallelism, request batching tends to densify expert activation in both the prefill and decode stages.} The union of token-level routing decisions across requests can activate a large portion, or even all, experts in a layer. This behavior weakens the effectiveness of prefetch-based scheduling and introduces significant pressure on GPU memory capacity as well as host–device communication bandwidth.

\textbf{Challenge \#2: Design expert scheduling strategies for prefill and decode stages. } Handling the two different activation modes during the inference phase of LLMs presents a significant challenge. In the prefill stage, computing with all experts simultaneously is certainly the fastest option. In inference systems that do not consider memory limitations, such as vLLM, operator fusion is applied in MoE layers, which allows different experts to complete parallel computations within the same operator. However, this approach substantially increases memory consumption, which is unacceptable in resource-constrained scenarios. Thus, minimizing memory usage while reducing latency becomes the core challenge for prefetching scheduling strategies. For the decode stage, although only the activated experts for a single token need to be loaded, the efficiency is highly dependent on the accuracy of the prediction model. If the prediction model makes an error, reloading the correct experts from CPU memory into GPU memory can significantly increase inference latency. Therefore, while efforts should be made to improve prediction model accuracy, a solution is still needed for handling cases where predictions fail. This requires a balance between prefetching efficiency and corrective measures that minimize the latency impact of incorrect predictions. In addition, under single-GPU deployments without expert parallelism, request batching can significantly densify expert activation and reduce the sparsity benefits of MoE inference. Therefore, DuoServe-MoE focuses on single-request serving to preserve sparse expert execution and achieve stable QoS-oriented latency.
\section{System Overview}

\begin{figure}[t]
  \centering
  \includegraphics[width=0.45\textwidth]{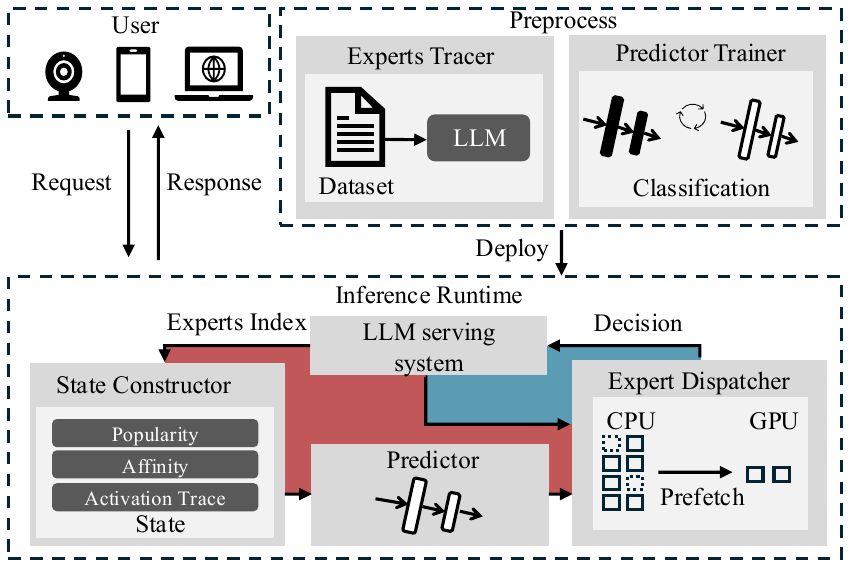} 
  \caption{System Overview}
  \label{promoe_arch}
\end{figure}

To address the two issues mentioned earlier, we propose an inference serving system for MoE-based LLMs called DuoServe-MoE. Figure \ref{promoe_arch} illustrates the system architecture and workflow of DuoServe-MoE, which is divided into two main stages: offline preprocess and online inference runtime.

In the offline preprocess stage, DuoServe-MoE profiles the expert activation trace using a dataset to create a simplified dataset. It captures the activation characteristics of the experts, including their popularity and inter-layer affinity. We use these data as inputs to the predictor, formulating the prediction of the next layer's activated experts as a classification task for training. Once training is completed, the model is deployed into the inference runtime. 

In the online phase, users send requests to DuoServe-MoE, which enters the inference runtime. In our scheduling design for the two phases of LLM inference, Figure \ref{promoe_arch} shows the blue area representing the prefill stage, while the red area represents the decode stage. Due to the dense activation of expert models during the prefill stage, the LLM serving system directly interacts with the Expert Dispatcher and employs a pipelining strategy to parallelize computation and communication.  In the decoding stage, the inference runtime sends the experts index of the activated experts to the State Constructor. The State Constructor integrates the activated experts with the popularity and affinity information collected during the preprocess stage. This consolidated data is then passed to the trained predictor from the preprocessing stage for predicting expert activations in subsequent layers. The Expert Dispatcher uses all of the predictor’s results to deploy the corresponding expert weights onto the GPU before the runtime moves on to the next MoE layer. Once all inference computations are completed, DuoServe-MoE sends the final result back to the user. This two-stage design allows DuoServe-MoE to reduce memory footprint while ensuring minimal inference latency, optimizing both the prefill and decode stages to meet the constraints of resource-limited environments effectively.

\section{Preprocess}
To effectively solve the problem of predicting multiple expert selections, we designed a learning-based method to train a predictor in the Prepocess stage of DuoServe-MoE. As discussed in Section \ref{2.1}, fine-tuning a Large Language Model (LLM) for expert prediction poses significant challenges in resource-constrained environments. Additionally, expert predictions depend heavily on inter-affinity and prior expert selections, making iterative predictions particularly detrimental to performance—especially in LLMs with numerous layers. Considering these factors, we propose a layer-level prediction method. Moreover, the data collection, training of the predictor, and final deployment of the LLM can all be accomplished on the same device, thereby reducing additional costs for users and making the solution more accessible for resource-limited scenarios.

\subsection{Data Collection}\label{4.1}
Based on the analysis of Section \ref{2.1}, the factors of expert selection can be formulated as three: experts' popularity, inter-layer affinity, and the past expert's activation path. Therefore, the Preprocess starts from the data collection. Experts Tracer utilizes a small portion of the dataset to record the expert activation paths during the inference process of LLMs. Each inference instance involves the selection of multiple experts from a set of available experts at each layer. We define an “expert activation path” as the sequence of selected experts across all layers during a single inference episode. For each layer $l$, the set of selected experts $E_l$ is recorded, leading to the following representation of activation paths across $L$ layers:
\begin{equation} E = \{E_l \mid l \in [1, L], E_l \subseteq \{e_1, e_2, \dots, e_M\}\}, \end{equation}
where $E_l$ represents the set of experts selected at layer $l$, and $M$ is the total number of available experts. This data forms the basis for constructing both the popularity and affinity matrices, which provide insight into expert selection patterns and inter-layer dependencies.

The popularity matrix $P$ at layer $l$ captures the frequency with which each expert $e_i$ is selected. To construct this matrix, we count the number of times each expert is chosen across all inference episodes. The popularity of expert $e_i$ at layer $l$ is calculated as:
\begin{equation}
P_l(i) = \frac{\sum_{n=1}^{N} \mathbbm{1}(e_{i} \in E_{l,n})}{\sum_{m=1}^{M} \sum_{n=1}^{N} \mathbbm{1}(e_{m} \in E_{l,n})},
\end{equation}
where $\mathbbm{1}$ is an indicator function that evaluates to 1 if expert $e_i$ is selected in the set $E_l$ for the $n$-th episode. After counting, the popularity matrix is normalized across each layer to represent the probability distribution of expert selections.

At last, the affinity matrix $A$ captures the relationship between expert selections in consecutive layers. For any expert $e_i$ selected in layer $l$, the matrix entry $A_{l,l+1}(i,j)$ represents the likelihood of selecting expert $e_j$ in layer $l+1$, given that $e_i$ was selected in layer $l$. This is computed by examining the joint selection of experts across consecutive layers:
\begin{equation}
A_{l,l+1}(i,j) = \frac{\sum_{n=1}^{N} \mathbbm{1}(e_{i} \in E_{l,n} \wedge e_{j} \in E_{l+1,n})}{\sum_{m=1}^{M} \sum_{n=1}^{N} \mathbbm{1}(e_{i} \in E_{l,n} \wedge e_{m} \in E_{l+1,n})}.
\end{equation}
Similar to the popularity matrix, the affinity matrix is normalized so that the sum of probabilities for selecting any expert in layer $l+1$, given a selection in layer $l$, equals 1. This matrix helps to identify inter-layer expert selection dependencies.

\subsection{Predictor Training Process}\label{4.2}
\textbf{Input Construction.} The task of predicting expert selections for the next layer is modeled as a supervised learning problem. Given the experts selected from all previous layers, along with the computed popularity and affinity matrices, the goal is to predict the set of experts that will be selected in the next layer. Let $h_l$ represent the set of all selected experts from the previous layers up to layer $l-1$:
\begin{equation}
h_l = \{E_0, E_1, \dots, E_{l-1}\},
\end{equation}
The input to the model is constructed by combining this set $h_l$, along with the popularity information $p_l$ for layer $l$, and the affinity matrix values $a_{l-1,l}$ between layers $l-1$ and $l$:
\begin{equation}
s_l = [h_l, p_l, a_{l-1,l}].
\end{equation}
To handle varying lengths in $h_l$, we flatten the expert indices and apply zero-padding to a fixed length. The affinity matrix $a_{l-1,l}$ and popularity vector $p_l$ are similarly flattened. All components are concatenated into a fixed-length input vector for the MLP. By using the cumulative set of all previous expert selections, the model is able to capture historical selection patterns, which improves its predictive ability for the next layer.

\textbf{Model Architecture and Training.} To predict the expert selections for the next layer, we design a deep seven-layer Multi-Layer Perceptron (MLP) named ExpertMLP, which takes the constructed state vector $s_l$ as input and outputs a multi-label probability distribution over all experts in the target layer. Compared to traditional shallow MLPs, our model introduces increased representational capacity to capture the nuanced patterns in expert activation conditioned on popularity, inter-layer affinity, and historical selection paths.

The architecture consists of seven fully connected layers with intermediate hidden dimensions progressively reduced from 2048 to 64, followed by a final output layer whose dimension matches the number of experts. Each layer is followed by a Batch Normalization layer and ReLU activation, along with Dropout (with a default rate of 0.1) applied after each hidden layer to prevent overfitting. This design balances depth and regularization, enabling the model to generalize well across varying expert selection distributions.

Because DuoServe-MoE needs to predict multiple activated experts, which can be considered as multiple labels. Therefore, the model is trained using a Binary Cross-Entropy loss function\cite{ruby2020binary}, which compares the predicted probabilities of the experts with the ground truth expert selections. The loss function is defined as:
\begin{align}
L(\theta) = - \sum_{n=1}^{N} & \left[ y_n \log(\sigma(f_\theta(s_n))) \right. \nonumber \\
& \left. + (1 - y_n) \log(1 - \sigma(f_\theta(s_n))) \right],
\end{align}
where $y_n$ is the one-hot encoded ground truth expert selection for the $n$-th sample, and $\sigma$ is the sigmoid function used to convert the model’s outputs to probabilities.

This training process allows the model to learn the expert selection patterns using the previously computed popularity and affinity matrix, leveraging the relationships across multiple layers to improve predictive accuracy.

\section{Inference Runtime}
\begin{figure*}[t]
    \centering
    \subfigure[Prefill]{
        \includegraphics[width=0.45\linewidth]{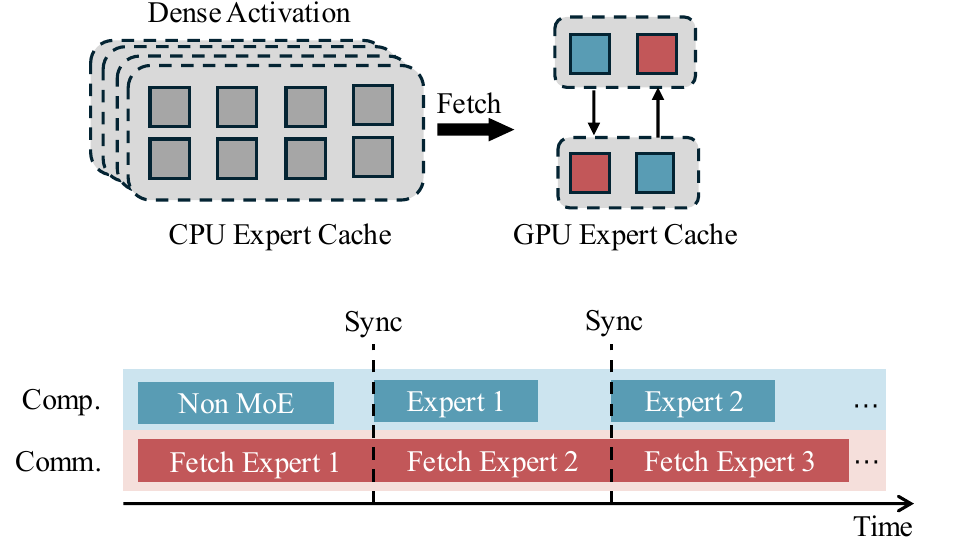}
        \label{fig:prefill}
    }
    \subfigure[Decode]{
        \includegraphics[width=0.48\linewidth]{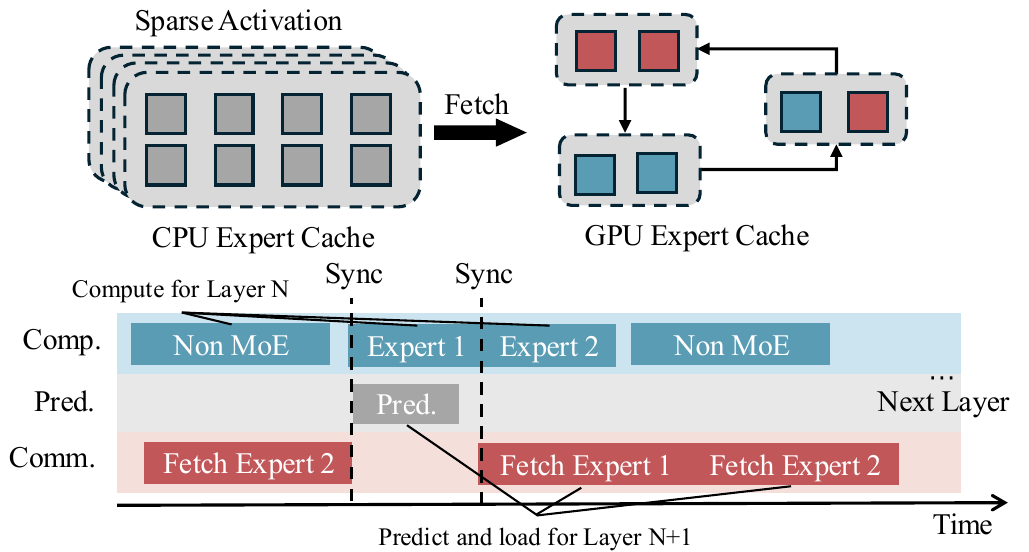}
        \label{fig:decode}
    }
    \caption{Prefill and Decode scheduling mechanism in DuoServe-Moe. The top side of the figures is the memory view, and the down side represents the CUDA streams view. }
    \label{fig:prefill and decode}
\end{figure*}
\subsection{Inference Runtime Workflow}
At the start of the inference runtime, DuoServe-MoE fetches the model weights onto the GPU. In MoE-based LLMs, the non-MoE model weights account for only about 10\% of the total weights, so DuoServe-MoE chooses to fetch them all onto the GPU to maintain sufficient inference performance. For the MoE layer weights, DuoServe-MoE establishes an expert cache on the GPU. Weights for activated experts are prefetched from the CPU expert cache to the GPU expert cache just before expert execution.

To control memory overhead, DuoServe-MoE keeps only a small number of experts resident on GPU at runtime. In our single-request decode setting, the GPU expert cache is sized to match the per-token activated expert count. Therefore, for MoE models with Top-$k$ routing, we set the GPU expert cache size to $k$. For example, in Mixtral 8$\times$7B with Top-2 routing, the cache size is set to 2. This design keeps GPU residency lightweight while still allowing expert prefetching and computation to overlap through the dual-stream scheduling mechanism.

Once the serving system has deployed the model, users send requests to the inference runtime, and the Expert Dispatcher manages the prefetching and scheduling of the experts. Due to the different activation modes of expert models in the prefill and decode stages, we employ two distinct scheduling approaches. Figure \ref{fig:prefill and decode} illustrates DuoServe-MoE's mechanism for handling expert layers from the perspectives of memory scheduling and CUDA streams. In the prefill stage, given the large number of tokens and all experts being activated, there is no need for predictions from the predictor. We use the Mixtral model as a demonstration case, where each layer has 8 experts, and 2 of them are selected for activation. Therefore, the CPU expert cache size is 8 experts per layer, while the GPU expert cache size is 2. If DuoServe-MoE serves other LLMs, the size of the expert cache will be according to the model architecture. 

In the Expert Dispatcher, we utilize a multi-CUDA stream\cite{cudastream} pipeline technique, allowing weight communication and operator computation to be executed concurrently, which can minimize latency and keep memory consumption low. In the decode stage, we leverage the predictor trained during the preprocess phase to predict future experts, thereby reducing additional communication overhead. 

\subsection{Prefill Stage}
As shown in Figure \ref{promoe_arch}, the LLM serving system directly manages scheduling through the Expert Dispatcher during the prefill stage. Since all experts need to be processed, we focus on how to fetch all the experts of each layer into the GPU expert cache and compute them efficiently. 

We establish two CUDA streams: one as the computation stream and the other as the communication stream. DuoServe-MoE uses the communication stream to prefetch the first expert weights into the GPU expert cache, while the computation stream concurrently performs computation for the non-MoE layers. Before entering the MoE layer, we synchronize the two streams to ensure that the expert weights are loaded into the GPU expert cache. This approach minimizes latency by overlapping communication and computation, thus maximizing the leverage of GPU resources during the prefill stage.

After the gate function in the MoE layer calculates the expert selection for all tokens, the LLM serving system groups the tokens by their assigned experts, ensuring that each expert performs a batch-processing operation for all associated tokens. Therefore, each expert weight only needs to be fetched once, optimizing memory operations. The system enters a two-stage pipelined phase. The communication stream starts fetching the next expert from the CPU expert cache once the first expert in the GPU expert cache has completed the computation. Due to the high parallelism of GPU, the operator in the computation stream is completed fast. However, constrained by the limited PCIe bandwidth, fetching expert weights in the communication stream is slower compared to the expert operator computation. As shown in the stream view in Figure \ref{fig:prefill}, we set synchronization points after each expert is fetched to ensure that no incorrect expert weights are used for computation. In the memory view of Figure \ref{fig:decode}, the two experts in the GPU expert cache are always in a state where one is being used for computation and the other is being fetched. This ensures that the communication stream is constantly fetching new weights without having to wait for the computation stream, thereby minimizing GPU memory usage while ensuring low latency. This balance between prefetching and computation effectively optimizes resource utilization during the prefill phase. Once all experts have completed their computations, the LLM serving system unpacks the results and finalizes the computation for each individual token. After all layers have completed their computations, the LLM generates the first token and proceeds to the decoding stage. 
\subsection{Decode Stage}
In the inference phase of an LLM, multiple iterations of computation are required to complete a request. Only the first round is the prefill stage, while all subsequent rounds are in the decoding stage, making optimization of the decoding stage crucial. Unlike the prefill stage, the decoding stage only processes one token at a time. This means that only 2 experts per layer need to be loaded. As illustrated in Figure \ref{promoe_arch}, during the decoding stage, we utilize the prediction model trained in the preprocessing phase to predict the expert selection for the current token. 

In the first layer, the Expert Dispatcher fetches the expert models into the GPU after the gate function completes the selection. Simultaneously, the LLM serving system sends the first layer's expert selection to the State Constructor for storage. The State Constructor retrieves the corresponding popularity vector for the upcoming layer and combines it with the inter-layer affinity vector from the previous layer's expert selection, using this data as input for the predictor. Since the predictor itself is also a neural network, which involves the computation on GPU, we add an additional prediction stream on top of the prefill stage to achieve parallel processing and minimize the extra overhead of prediction. As shown in Figure \ref{fig:decode}, when Layer N begins the expert computation, the predictor starts predicting the next layer's experts. After the first expert in the GPU expert cache completes its computation, the communication stream immediately starts prefetching the experts for the next layer.

To prevent race conditions during parallel operations, DuoServe-MoE has two synchronization points during the decoding stage. The first synchronization point ensures that prefetching has finished before Expert 1 begins computation, given that prefetching latency is higher than the operator computation in the computation stream. Additionally, the experts selected by the gate function are compared with those loaded in the GPU expert cache. If there is a mismatch, the correct experts are re-fetched from the CPU expert cache to ensure that the expert weights used are correct. The second synchronization point occurs after the first expert completes its computation and after the predictor in the prediction stream has predicted the next layer's expert selection, allowing prefetching for the new layer to begin. After each round of computation, the State Constructor clears the stored activation trace and incorporates the new expert selections for the next round, continuing this process until the request is fully processed.

\section{Evaluation}
\subsection{Experimental Setup}
\textbf{Implementation.} To implement DuoServe-MoE, we utilized the state-of-the-art LLM inference framework vLLM and employed CUDA pinned memory\cite{cuda} to enable efficient CPU offloading. Additionally, we leveraged CUDA streams to implement a multi-stream pipeline design, enhancing parallelism and improving system efficiency. For the prediction model, we used PyTorch for training and deployed it on the GPU to minimize latency. 
%统计PCIE的带宽

\textbf{Hardware.} Since DuoServe-MoE's performance is influenced by different scenarios, we evaluated it under two distinct settings in our experiments. The first scenario involves a GPU-based edge server, which consists of an AMD Threadripper PRO 3945WX 12-Core CPU, 128GB RAM, and an NVIDIA RTX A5000 GPU with 24 GB GPU memory, running on Ubuntu 20.04 OS. The second scenario involves using a more powerful GPU-based edge server, specifically an AMD Ryzen Threadripper PRO 5955WX 16-Core CPU, 128GB RAM, and an NVIDIA RTX A6000 GPU with 48 GB GPU memory, running on Ubuntu 24.04 OS. Both of them use PCIe4.0 with 16GT/s bandwidth. This scenario allows us to evaluate DuoServe-MoE's scalability across different computational resources by comparing its performance.
\begin{table}[t]
\caption{Major configurations of the backbone models.}
\label{table1}
\centering
\footnotesize
\setlength{\tabcolsep}{3pt}
\renewcommand{\arraystretch}{1.1}
\begin{tabular}{lccc}
\toprule
Model & Layers & Experts (Act./Tot.) & Params (Act./Tot.) \\
\midrule
Mixtral-8x7B & 32 & 2/8 & 12.9B/46.7B \\
Mixtral-8x22B & 56 & 2/8 & 39B/141B \\
Qwen3-30B-A3B & 48 & 8/128 & 3B/30B \\
DeepSeekMoE-16B & 28 & 8/66 & 2.8B/16.4B \\
\bottomrule
\end{tabular}
\end{table}

\textbf{Models.} We evaluated DuoServe-MoE using two different sizes of Mixtral, Qwen3-30B-A3B, and DeepSeekMoE-16B, all of which are decoder-only LLMs. In our single-GPU setting, both CPU memory and GPU memory are constrained. Therefore, we use 4-bit AWQ\cite{lin2024awq} for Mixtral, deploy an FP8-quantized version for Qwen3-30B-A3B, and run inference with vLLM quantization operators. For DeepSeekMoE-16B, because the total weight size is 16B, we deploy the full model weights. We downloaded all pre-trained weights from Hugging Face. Table \ref{table1} summarizes the model configurations. Both Mixtral variants use 8 total experts per MoE layer with top-2 routing (2 activated experts per token), Qwen3-30B-A3B uses 128 total experts with 8 activated experts per token, and DeepSeekMoE-16B uses 66 total experts with 8 activated experts per token.

\textbf{Workload.} In our main experiments, we evaluate DuoServe-MoE under single-request scenarios, which reflect the target deployment environment of resource-constrained edge or single-GPU servers for latency-sensitive MoE serving. We further include a batching-throughput study as an extension experiment to examine how the proposed design behaves when multiple requests are processed together.

Since batching introduces different expert activation patterns during decoding, single-request evaluation remains our primary setting, while batching is included as an extension study to further examine the robustness and scalability of DuoServe-MoE under more practical serving workloads. We selected two datasets to evaluate DuoServe-MoE: the question-answering dataset SQuAD~\cite{rajpurkar2018know}, and Orca-Math~\cite{mitra2024orca}, a dataset focusing on mathematical reasoning tasks.

\textbf{Metrics.} For performance evaluation, we use standard LLM serving metrics, including \textbf{average end-to-end latency}, \textbf{time to first token (TTFT)}, and \textbf{total tokens per second under different batch sizes}. To further reflect request-level QoS, we additionally report \textbf{tail latency} in representative settings. We also analyze \textbf{peak GPU memory usage} and the \textbf{prediction model hit rate}.

\textbf{Baselines.}
To analyze DuoServe-MoE's (\textbf{DuoServe}) performance, we compared it against three method-oriented baselines. All baselines offload non-activated weights to CPU memory to reduce GPU memory usage. The first baseline is \textbf{On-Demand Fetch} (\textbf{ODF}), which loads activated experts onto GPU only after gate selection (implemented with Hugging Face Accelerate\cite{huggingface2022accelerate}). The second baseline is \textbf{Layer-wise Full Prefetch} (\textbf{LFP}), which prefetches all experts of each layer into GPU memory before expert computation (following MoESys\cite{yu2024moesys}). The third baseline is \textbf{MoE-Infinity} (\textbf{MIF}), which exploits request-level expert activation tracing to guide activation-aware expert prefetching and caching for efficient MoE offloading\cite{xue2024moe}.
\begin{figure*}[t]
  \centering
  \includegraphics[width=0.92\textwidth]{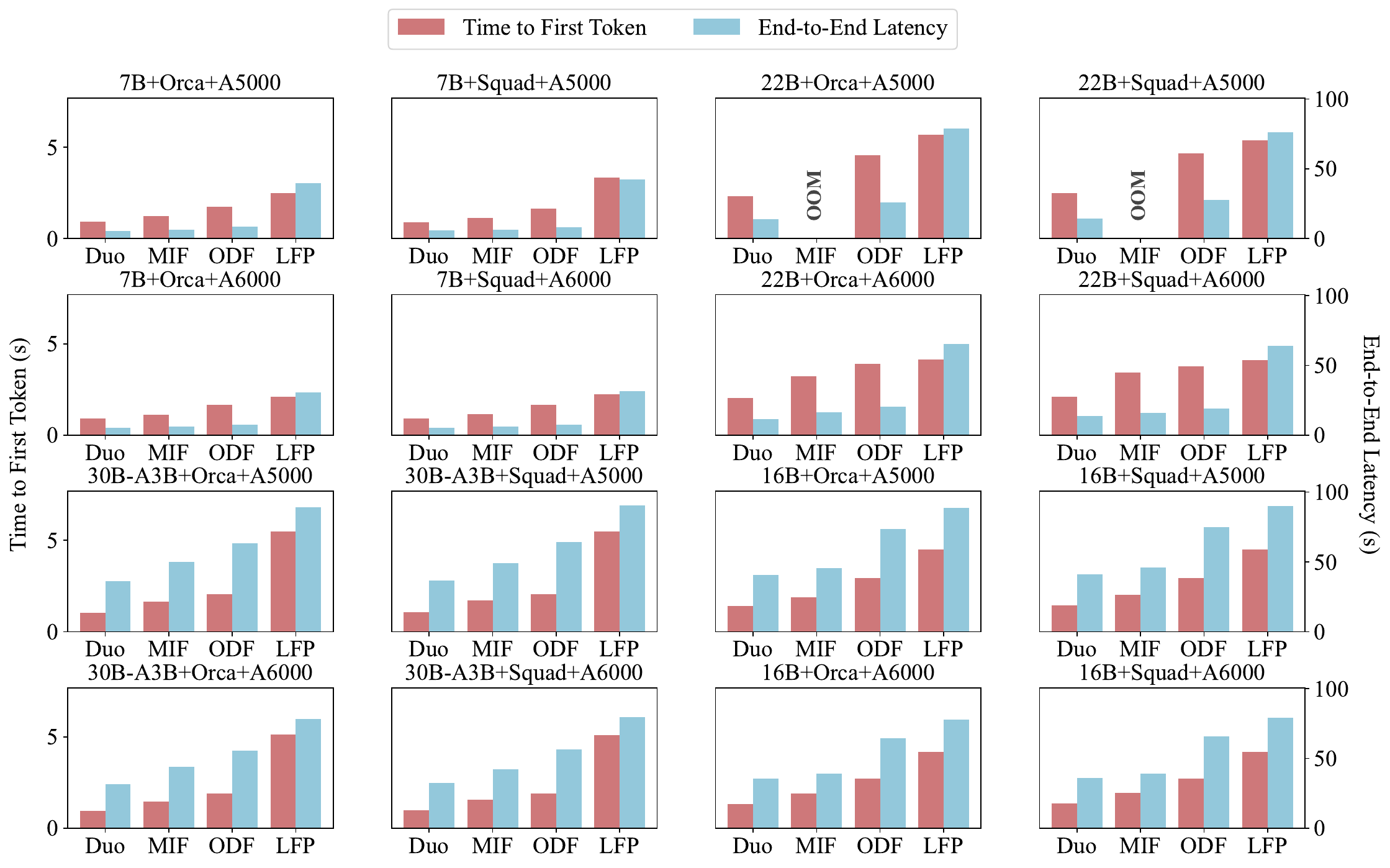} 
\caption{Average TTFT and end-to-end latency across different models, datasets, and hardware platforms.}
  \label{Bar}
\end{figure*}

\subsection{Performance Evaluation}
\begin{figure}[t]
    \centering
    \includegraphics[width=0.92\linewidth]{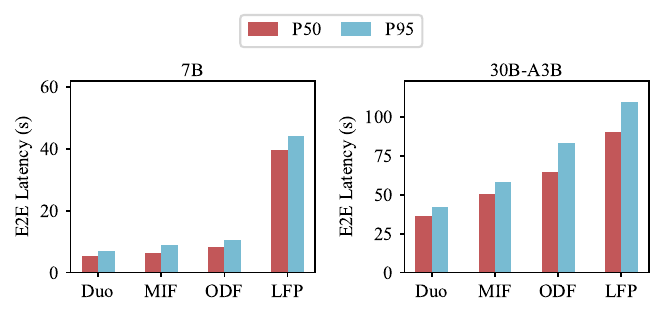}
    \caption{Tail latency comparison in representative settings on the A5000 GPU with the SQuAD dataset. We report the P50 and P95 end-to-end latency for two representative models: Mixtral-8x7B and Qwen3-30B-A3B.}
    \label{tail_latency}
\end{figure}

\textbf{Average Latency.} Figure~\ref{Bar} shows the average TTFT and average end-to-end latency across different models, datasets, and hardware platforms. Overall, DuoServe-MoE achieved an improvement of 1.78$\times$ to 5.34$\times$ in TTFT compared to ODF and LFP, while the end-to-end latency improved by 1.42$\times$ to 7.55$\times$. Compared to DuoServe-MoE and ODF, LFP exhibits much higher latency on the 7B and 22B models, primarily because LFP requires prefetching a full layer of expert models, which leads to excessive communication overhead. In contrast, for the more sparse 30B-A3B and 16B models, ODF becomes less competitive, since loading more activated experts on demand exposes higher transfer latency on the critical path. MIF improves over ODF and LFP in some settings by using activation-aware tracing and prefetching, but it still remains consistently slower than DuoServe-MoE when it is available, and it encounters OOM on the 22B model on A5000. This indicates that DuoServe-MoE's phase-specialized design is more effective at overlapping communication with computation. In particular, DuoServe-MoE achieves over 5$\times$ TTFT improvement over LFP on the 30B-A3B model, while its TTFT improvement over ODF on this model is about 1.95$\times$ to 1.99$\times$. Compared to the baselines, DuoServe-MoE benefits from phase-specialized optimization: during decode, the prediction model enables expert prefetching and overlaps communication with computation, while during prefill, DuoServe-MoE performs computation on one expert while prefetching another, thereby reducing latency through pipelining.

When comparing datasets, DuoServe-MoE performs slightly better on Orca, achieving an average 1.72$\times$ improvement in end-to-end latency over ODF, while on SQuAD the improvement is 1.67$\times$. Although the absolute latency on A6000 is lower than that on A5000, DuoServe-MoE maintains similar average TTFT improvements over ODF and LFP on both platforms, showing that the proposed method remains effective under different hardware capabilities. MIF follows the same general trend across devices, but its advantage is less stable and its higher memory cost further limits practicality. Overall, DuoServe-MoE maintains strong and scalable latency benefits for practical MoE serving.

\textbf{Tail Latency.} In addition to average latency, we further evaluate request-level QoS by reporting the P50 and P95 end-to-end latency in representative settings, as shown in Figure~\ref{tail_latency}. DuoServe-MoE consistently achieves lower tail latency than ODF and LFP on both models, indicating that the proposed phase-specific scheduling not only improves mean performance but also mitigates high-latency cases caused by expert transfer stalls. MIF also reduces tail latency in several cases, but it still shows higher P50 and P95 latency than DuoServe-MoE, suggesting that its trace-based mechanism is less effective in stabilizing latency under dynamic request patterns. We select Mixtral-8x7B and Qwen3-30B-A3B as two representative models because they capture two distinct sparsity regimes in our evaluation: Mixtral-8x7B represents the Mixtral-style setting with a relatively small expert pool and top-2 routing, while Qwen3-30B-A3B represents a much sparser setting with a larger expert pool and top-8 routing. Since these two cases already cover the main latency behaviors observed in our experiments, we use them as concise representatives to keep the tail-latency evaluation lightweight and space-efficient. The results show that DuoServe-MoE improves not only average performance but also latency stability under resource-constrained single-GPU MoE serving.
\begin{figure}[t]
    \centering
    \includegraphics[width=0.92\linewidth]{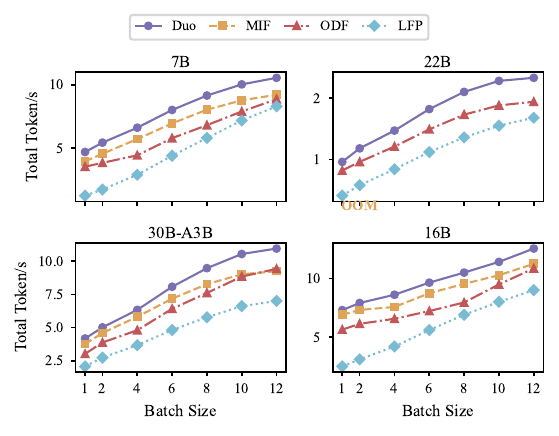}
    \caption{Total throughput under different batch sizes on A5000 using the SQuAD dataset. We evaluate four MoE models with batch sizes ranging from 1 to 12.}
    \label{throughput_bs}
\end{figure}

\textbf{Batching Throughput.} To further understand how DuoServe-MoE behaves beyond the single-request setting, we extend the evaluation to batched inference and report the total throughput under different batch sizes, as shown in Figure~\ref{throughput_bs}. Overall, the throughput of all methods increases as the batch size grows, since more requests can share the non-MoE computation and amortize part of the runtime overhead. DuoServe-MoE consistently achieves the highest throughput across all evaluated models, while MIF remains the second-best method when it is available and still encounters OOM on Mixtral-8x22B.

The advantage of DuoServe-MoE under batching comes from the same phase-specialized design used in the single-request setting. During prefill, DuoServe-MoE overlaps expert prefetching with non-MoE computation through a two-stream pipeline, which continues to reduce the exposed communication cost when multiple requests are processed together. During decode, although batching increases the union of activated experts and weakens sparsity, DuoServe-MoE still benefits from prediction-guided prefetching and communication-computation overlap, so a substantial part of the transfer latency can still be hidden. In contrast, ODF places expert transfer more directly on the critical path, while LFP suffers from excessive communication and memory overhead due to aggressive full-layer prefetching. MIF improves over ODF and LFP by using trace-based activation-aware prefetching, but its benefit is less stable under larger batches because routing becomes more diverse and its tracing-based mechanism is less adaptive than DuoServe-MoE's learned predictor.

We also observe that the throughput gain becomes less pronounced as the batch size continues to increase, especially on larger and sparser models. This is consistent with our motivation analysis: batching densifies expert activation, enlarges the union of experts within a layer, and reduces the sparsity benefit that DuoServe-MoE exploits during decoding. Even under this less favorable setting, DuoServe-MoE still maintains the best throughput among all compared methods, showing that its scheduling design remains effective and robust under moderate batching.
\subsection{Memory Consumption Analysis}
To evaluate memory efficiency, we measure peak GPU memory usage. Since device differences do not affect memory usage and dataset differences only introduce minor variation through KV cache size, Table~\ref{table2} reports memory statistics only across models. Besides the baselines and DuoServe-MoE, we also include a ``GPU Only'' setting for reference, where all model weights are placed on GPU. Although this setting provides the best possible performance, its memory cost is prohibitive. For example, Mixtral 8$\times$22B requires 138\,GB of GPU memory, which is infeasible in resource-constrained environments. Therefore, we do not treat GPU-only execution as a practical baseline.

Compared with the baselines, DuoServe-MoE uses less memory than LFP and MIF, while using slightly more memory than ODF. LFP incurs higher memory overhead because it places all experts of each layer on GPU, whereas ODF and DuoServe-MoE load only the required experts. MIF shows even higher memory consumption due to its tracing, prefetching, and caching mechanism, and it even runs out of memory on Mixtral-8x22B. Across all evaluated models, DuoServe-MoE consistently maintains a low GPU memory footprint while avoiding the excessive overhead of more aggressive prefetching-based methods. Overall, the results show that scheduling policy is the dominant factor in practical peak memory usage, and DuoServe-MoE achieves a favorable balance between memory consumption and latency.

\begin{table}[t]
\caption{Peak memory consumption across models. ``GPU only'' denotes full model deployment on GPU.}
\label{table2}
\centering
\footnotesize
\setlength{\tabcolsep}{3pt}
\renewcommand{\arraystretch}{1.1}
\begin{tabular}{lccccc}
\toprule
Model & LFP & ODF & MIF & DuoServe & GPU only \\
\midrule
Mixtral-8x7B & 4.24GB & 3.63GB &15.24GB & 3.91GB & 25.14GB \\
Mixtral-8x22B & 9.13GB & 8.06GB& OOM & 8.44GB & 138GB \\
Qwen3-30B-A3B & 1.80GB &  1.24GB&7.67GB & 1.54GB & 30.0GB \\
DeepSeekMoE-16B & 2.95GB & 1.98GB & 11.35GB & 2.28GB & 32.8GB \\
\bottomrule
\end{tabular}
\end{table}

\subsection{Predictor Overhead Discussion} \label{6.4}

In DuoServe-MoE, the additional overhead of the predictor mainly comes from data collection and training. To ensure that DuoServe-MoE can be fully deployed in resource-constrained environments, all stages should be executable on the same device, which is why we adopt a lightweight predictor. During the data collection phase (Section~\ref{4.1}), we select corresponding datasets for different tasks and use a scheduling strategy similar to ODF to ensure efficient trace collection. DuoServe-MoE uses only 2.5\% of the dataset for training, and the collection process takes about 5 to 8 hours depending on the device capability and model size. For training, the lightweight MLP-based predictor introduces limited additional cost, while models with larger expert pools and sparser routing patterns are relatively harder to learn due to the increased complexity of expert activation behavior. Nevertheless, across all four models, the entire preprocessing process, including both data collection and training, takes no more than 9 hours. Compared with fine-tuning an LLM, this overhead is modest and practical. Since the data requirement is small and the original LLM is not modified, users can also perform the preprocessing procedure alongside actual inference, allowing data collection and serving to proceed simultaneously without requiring a separate expensive preparation stage.

\begin{table}[t]
\caption{Prediction accuracy comparison between DuoServe and MIF.}
\label{table3}
\centering
\footnotesize
\setlength{\tabcolsep}{4pt}
\renewcommand{\arraystretch}{1.1}
\begin{tabular}{llcccc}
\toprule
\multirow{2}{*}{Model} & \multirow{2}{*}{Dataset} 
& \multicolumn{2}{c}{Top-k} 
& \multicolumn{2}{c}{At Least Half} \\
\cmidrule(lr){3-4} \cmidrule(lr){5-6}
& & DuoServe & MIF & DuoServe & MIF \\
\midrule
\multirow{2}{*}{Mixtral-8x7B}
& Orca  & \textbf{66.85\%} & 42.33\% & \textbf{95.45\%} & 81.32\% \\
& Squad & \textbf{60.23\%} & 40.13\% & \textbf{92.31\%} & 80.22\% \\
\midrule
\multirow{2}{*}{Mixtral-8x22B}
& Orca  & \textbf{56.21\%} & 38.22\% & \textbf{90.45\%} & 77.19\% \\
& Squad & \textbf{54.16\%} & 36.13\% & \textbf{90.31\%} & 76.83\% \\
\midrule
\multirow{2}{*}{Qwen3-30B-A3B}
& Orca  & \textbf{56.12\%} & 38.45\% & \textbf{98.14\%} & 82.33\% \\
& Squad & \textbf{55.33\%} & 38.22\% & \textbf{98.10\%} & 81.59\% \\
\midrule
\multirow{2}{*}{DeepSeekMoE-16B}
& Orca  & \textbf{60.24\%} & 44.33\% & \textbf{99.1\%}  & 82.13\% \\
& Squad & \textbf{61.33\%} & 42.14\% & \textbf{98.42\%} & 84.22\% \\
\bottomrule
\end{tabular}
\end{table}

Table~\ref{table3} further compares the prediction accuracy of DuoServe-MoE and MIF across different models and datasets. The Top-$k$ metric measures whether all routed experts are correctly predicted, while ``at least half'' measures whether at least half of the routed experts are correctly predicted. DuoServe-MoE consistently outperforms MIF on both metrics across all evaluated settings, showing that its lightweight predictor captures expert routing behavior more accurately. This is mainly because MIF relies on trace-based activation matching, which is less effective when expert routing varies across requests, whereas DuoServe-MoE learns routing patterns directly and can better adapt to such variation. Although prediction becomes slightly harder on larger models, the overall trend remains stable across datasets. In runtime, the predictor uses only about 300\,MB of GPU memory and introduces around 0.6\,ms latency, which can be further hidden by CUDA-stream parallelization. Therefore, the overhead of the predictor is small and acceptable in practice.
\section{Related Work}
\textbf{Efficient LLMs inference serving.} With the rapid development of Large Language Models, efficient inference serving has become a key area of research. Some prior work focuses on optimizing models through compression \cite{frantar2024qmoe} techniques, such as model pruning \cite{chen2022task,frantar2023sparsegpt,sun2024a}, knowledge distillation \cite{fu2023specializing,wu-etal-2024-lamini}, and quantization \cite{lin2024awq,frantar2023optq}. These methods can be combined with DuoServe-MoE, which uses AWQ quantization for model deployment. Additionally, some methods address the bottleneck of insufficient GPU memory by using CPU offloading\cite{sheng2023flexgen,xuanlei2024hetegen,ren2021zero} and optimizing the KV cache \cite{kwon2023efficient,adnan2024keyformer}. These methods are better suited for dense models. DuoServe-MoE also uses CPU offloading but is designed specifically with GPU efficiency in mind.

\textbf{Prefill and decode optimization methods.}
Another line of work studies the heterogeneity between prefill and decode in autoregressive inference. Sarathi-Serve proposes chunked prefill to mitigate head-of-line blocking~\cite{298679}, while FastServe uses token-level preemption to reduce latency tails~\cite{wu2023fast}. DistServe disaggregates prefill and decode across GPUs~\cite{298687}, and Nexus further explores intra-GPU disaggregation to reduce coordination overhead~\cite{shi2025nexus}. POD-Attention, TaiChi, and FlashDecoding++ improve phase-specific efficiency through adaptive kernels, hybrid scheduling, and asynchronous decoding pipelines~\cite{kamath2025pod,zhou2025taichi,hong2024flashdecoding++}. These studies show that prefill and decode exhibit substantially different performance characteristics and should not be optimized with a unified policy. However, they mainly focus on dense LLM serving or multi-request scheduling, and do not explicitly address the phase-specific expert scheduling problem in resource-constrained single-GPU MoE inference.

\textbf{MoE Inference serving systems.}
For inference serving of MoE-based LLMs, existing studies can be broadly divided into two categories. The first focuses on distributed inference across multiple GPUs~\cite{yao2024exploiting,aminabadi2022deepspeed,li2023accelerating,shi2024schemoe}, where the main challenge lies in parallelizing expert computation and reducing inter-device communication through expert placement and scheduling strategies. The second category targets resource-constrained scenarios with a single GPU. These approaches~\cite{yi2023edgemoe,du2024sida,xue2024moe,kong2024swapmoe} generally rely on CPU offloading as the main scheduling principle, often combined with prediction mechanisms for expert activation. More recent studies further extend this line in different directions. Yu et al.~propose fine-grained expert offloading guided by expert selection patterns and prompt semantics to improve the latency--memory trade-off in MoE serving~\cite{yu2025taminglatencymemorytradeoffmoebased}. KTransformers studies CPU/GPU heterogeneous inference for large MoE models through specialized CPU kernels and asynchronous task scheduling~\cite{chen2025ktransformers}. Expert-as-a-Service (EaaS) targets large-scale MoE deployment by disaggregating experts into independent stateless services to improve scalability, elasticity, and robustness~\cite{liu2025expert}. However, some methods~\cite{hwang2024pre,du2024sida} require fine-tuning of LLMs, while others emphasize heterogeneous execution or distributed expert services. In contrast, our work focuses on resource-constrained single-GPU MoE serving and explicitly addresses the phase-specific expert scheduling problem caused by the different expert activation characteristics in the prefill and decode stages.
\section{Conclusion}
This paper presents DuoServe-MoE, an inference serving system designed to address two key challenges in deploying MoE-based LLMs in resource-constrained service environments. DuoServe-MoE employs two phase-specific expert scheduling mechanisms to handle the distinct expert activation patterns in the prefill and decode stages. Additionally, it leverages a lightweight learning-based predictor to prefetch likely activated experts, reducing communication overhead while maintaining low GPU memory usage. Experimental results show that DuoServe-MoE achieves 1.78$\times$ to 5.34$\times$ improvement in TTFT and 1.42$\times$ to 7.55$\times$ improvement in end-to-end latency, while maintaining low runtime GPU memory consumption. These results demonstrate the effectiveness of DuoServe-MoE for QoS-oriented MoE LLM serving in resource-constrained single-GPU deployments.

\bibliographystyle{IEEEtran} 
\bibliography{references}

\end{document}